\documentclass[sigconf]{acmart}
\usepackage{svg}

\AtBeginDocument{%
  }

\setcopyright{none}
\acmDOI{}
\acmISBN{}
\acmYear{}
\acmConference[LOCO '26]{Make sure to enter the correct
  conference title from your rights confirmation email}{September 10--11,
  2026}{Lancaster, UK}

\begin{document}

\title{The ultimate carbon cost of a ChatGPT query}

\author{Paul Kron}
\email{contact@mailpk.eu}

\renewcommand{\shortauthors}{Kron}

\begin{abstract}
This paper reviews and combines findings from the fields of product and life-cycle analysis \cite{schneider_et_al_life-cycle_2025, you_how_2025}, the usage of modern transformer-based large language models (LLM) \cite{chiang_et_al_chatbot_2024}, as well as on greenhouse gas emissions and the ultimate cost of their subsequent consequences for future generations \cite{archer_et_al_ultimate_2020}. In this paper, it is shown that the carbon cost of a LLM query is in the order of magnitude of  $0.4 \frac{\$}{query}$ (USD) for the future human population in the form of environmental disruptions . This corresponds to emissions in the magnitude of $10 \frac{gCO_2eq}{query}$. The most significant unknown factor in that calculation being the number of tokens computed (1k to 100k tokens equal $1.2 \frac{cent}{query}$ to $120 \frac{cent}{query}$). 
This number is subject to a wide range of calculation uncertainties and is less to be seen as a matter of fact and more as an order of magnitude estimate. This estimate is aimed towards aiding the discourse surrounding AI systems by uncovering the inevitable consequences of technological development by the means of attaching a consequence in a familiar unit to it. By the introduction of the per query ultimate carbon cost (QCC), even if attached to great uncertainty, it is highlighted that the use of AI services happens within hypercomplex interdependent systems and has concrete consequences for our planetary health. 
The spread of the awareness about the interdependence of planetary health and AI usage can be useful for the individual user in the formation of political opinion through discourse \cite{brennan_considering_2017} as well as a literate usage of AI systems \cite{pinski_ai_2024}. Ways to increase the accuracy of the estimations, such as incorporating the cost of AIs water consumption or further determining the realistic token count of a query, have been identified as further research targets. 
\end{abstract}




\setcopyright{none}
\settopmatter{printacmref=false}
\acmDOI{}
\acmISBN{}
\acmYear{}

\maketitle

\begin{table}[h!]
  \caption{Abbreviations}
  \label{tab:abr}
  \begin{tabular}{ccc}
    \toprule
    Abbreviation & Name & Unit \\
    \midrule
CCI          & Carbon Compute Intensity                 & $\frac{gCO_2eq}{10^{18}FLOP}$ \\
 CI           & Grid Carbon Intensity                    & $\frac{gCO2eq}{kWh}$      \\
 CPU          & Central processing unit                  &                           \\
 FLOP         & Floating point operation                 &                           \\
 LLM          & Large Language Model                     &                           \\
 $M_{C}$/$M_{CO_2}$ & Molecular weight of carbon/carbon dioxide & 12/44 $\frac{g}{mol}$  \\
 QC           & Query required Compute                   & $\frac{FLOP}{query}$      \\
 QCC          & Query Carbon Cost                        & $\frac{\$}{query}$        \\
 QCI          & Query Carbon Intensity                   & $\frac{gCO_2eq}{query}$   \\
 TC           & Token required Compute                   & $\frac{FLOP}{Token}$      \\
 TCC          & Token Compute Cost                       & $\frac{\$}{Token}$        \\
 TCI          & Token Compute Intensity                  & $\frac{gCO_2eq}{Token}$   \\
 TPU          & Tensor processing unit                   &                           \\
 UCC          & Ultimate Carbon Cost                     & \$                        \\
 $c_{CO2u}$   & ultimate cost per ton of oxidized carbon & $\frac{\$}{tCO_2eq}$      \\
 $c_{Cu}$     & ultimate cost per ton of carbon          & $\frac{\$}{tC}$           \\
 $r_t$        & FLOP per Token compute ratio             & $\frac{FLOP}{Token}$      \\
 $t_q$        & query length                             & $\frac{Token}{query}$     \\
 $w_a$        & active Model size                        & weights                   \\
  \bottomrule
\end{tabular}
\end{table}

\section{Introduction} 

The marketing umbrella term of 'Artificial Intelligence' for a wide range of computing technology has become very dominant in societal and academic discourse over the past years with the arrival of a wide range of consumer services such as ChatGPT - a low friction, LLM-based chat interface. Thereby also many people with little to no technical expertise are confronted with the opportunity to use such services. Meanwhile the literacy on the subject, including the environmental implications, appear to lag behind. Gaining knowledge on the nature of AI systems, such as the required work for data labelling (often conducted under precarious labour conditions) or the environmental impact of large scale computing demands, can lead to more widespread AI literacy. 'AI literacy' meaning the users "proficiency in different subject areas of AI that enable purposeful, efficient, and ethical usage of AI technologies" \cite{pinski_ai_2024}.   

\subsection{Motivation}
For users, researchers, and developers alike it is important to make an effort to understand the risks of LLM technology, especially since the benefits and risks are not equally distributed. This is highlighted by Bender et al. \cite{bender_et_al_dangers_2021} in their 2021 paper 'On the Dangers of Stochastic Parrots', which raises the example for environmental racism of 800,000 people in Sudan being affected by flooding, a phenomenon heightened by climate crisis, while there was not even one large-scale language model trained for Dhivehi or Sudanese Arabic. The paper furthermore advocates for more considerations on "how research contributions shape the overall direction of the field." \cite{bender_et_al_dangers_2021} and suggest a re-alignment of research goals away from what AI could hypothetically achieve and onto "understanding how machines are achieving the tasks in question and how they will form part of socio-technical systems.". The literacy on the ultimate carbon cost of their use is seen as a relevant quality to LLMs role in socio-technological interactions. While there is already a plethora of research on the topic, published \cite{faiz_llmcarbon_2024, jacquet_et_al_cinergy_2026, halgamuge_emissions_2026, luccioni_et_al_estimating_2023, kirkpatrick_carbon_2023, de_vries_growing_2023, devadas_towards_2026} and ongoing, those papers do not adequately address recent developments, such as the high token usage of generative, agentic or file processing systems, or provide complex results, that are not aimed to investigate an average users emissions. It is therefore proposed, that a quantified value of consequences of user actions, the query carbon intensity (QCC), expressed in \$ (a unit that is relevant an familiar to us), can aid the discourse surrounding AI services. Additionally the units of query carbon intensity (QCI), token carbon intensity (TCI) and token carbon cost (TCC) are proposed as intermediate steps for comparability. Influential works, like "Atlas of AI" by Kate Crawford \cite{crawford_atlas_2021}, are showing and dissecting the structures behind AI technology in a way that actively aids public discourse around the benefits and limitations of AI as suggested by Bender et al. \cite{bender_et_al_dangers_2021} and help others in forming literacy. The environmental risks of AI are especially important, because the water and power demands of AI computing present an infrastructural need, that is contradictory to the required sufficiency in power usage to meet essential goals in the prevention of climate catastrophes \cite{shukla_et_al_climate_2022} (Policy Paper p. 15).  

This study reviews and then combines carbon compute intensity (CCI) of the required hardware and infrastructure, the computing demands for a single search query using ChatGPT and then finally the ultimate cost of the $CO_{2}$ equivalent ($CO_{2}eq$) emissions on a long-lived human population. That way the very abstract concepts of computing are not only broken down to their $CO_{2}eq$ impact on our atmosphere, but also the abstraction layer between atmospheric changes and the impact on a human population is broken down via a quantified conversion.         
While this lengthy conversion is happening within a system of interdependent variables and in a space of partially very scarce or speculative base-data, it needs to be emphasised that this conversion itself is supposed to provide a statement on the interdependence of variables. Meaning while the exact figure is prone to high variation, it can be used to show, that the use of services such as ChatGPT has an ultimate cost to our planetary health attached to it, that is usually hidden behind many layers of abstraction. The underlying question this paper poses is thereby not 'what is the carbon cost of a ChatGPT query', but 'what can we learn about the ecological and social structures and systems that AI is embedded in by going down the path of this end-to-end calculation?'. 

\subsection{Ultimate carbon cost}
As D. Brennan states in reference to Hannah Arendt "a common world is created through shared conversation about the nature of things. The fluidity of a political situation is generated through the motion of continued conversation which constantly updates the shared understanding of the world [...] " \cite{brennan_considering_2017}. In which sense the introduction of a previously not made mathematical bridge the paper of Archer et al. \cite{archer_et_al_ultimate_2020} introduces an updated understanding of the world by defining carbon emissions into the global environment as ultimate costs for the whole of the future human population. The ultimate cost of carbon in an estimate of the "potential ultimate cost of fossil-fuel carbon to a long-lived human population over a one million–year time scale" by calculating the "[m]onetary costs [...] converted to units of present-day dollars by multiplying the future damage fractions by the present-day global world production, and integrated through time" \cite{archer_et_al_ultimate_2020}. The ultimate costs of one ton (1t) of carbon oxidised and released into the atmosphere range from \$10k to \$750 with a best-estimate value of \$100k, which is the value used later on.  

The carbon footprint of a LLM based system comes down to two separate values \cite{faiz_llmcarbon_2024}. On one side there is the operational footprint of an LLM, stemming from the raw hardware energy consumption. On the other side the embodied footprint encompasses emissions arising from the hardware production required to power the query. Those values are added together to reach the CCI as stated in the following. The footprint of the training is also divided by the estimated amount of queries to give an idea of the significance of its impact. This paper takes OpenAI's "ChatGPT" service as an example, more closely the GPT-4 model and its life-cycle within ChatGPT.  

\section{Calculations}
\subsection{Embodied emissions}
\subsubsection{Carbon Compute Intensity}\label{sec:CCI}

The cradle-to-grave based approach of Schneider et al. \cite{schneider_et_al_life-cycle_2025} (for Google) shows a total carbon footprint of 386 to 1,101 $\frac{gCO_2eq}{10^{18}FLOP}$(with a median of 788 $\frac{gCO_2eq}{10^{18}FLOP}$) for a 6 year life-cycle of different tensor processing unit (TPU) systems, as they would be used in AI training and processing. This includes TPU, CPU, Motherboard and Datacenter construction and is the sum of 38 to 114 $\frac{gCO_2eq}{10^{18}FLOP}$ of embodied and 316 to 929 $\frac{gCO_2eq}{10^{18}FLOP}$ of (location-based) operational carbon compute intencity (CCI). 
This is higher than estimations by A. d'Orgeval et al. that found CCI values between 60 and 214 $\frac{gCO_2eq}{10^{18}FLOP}$ \cite{dorgeval_et_al_generative_2026} while examining the life cycles of six different data centre architectures used for AI computing.   
A recent study by Falk et al. \cite{falk_et_al_more_2025} finds 0.0054 $\frac{kgCO_2eq}{h}$ in its multi-criteria cradle-to-grave analysis of an Nvidia A100 SXM 40 GB 3 year life-cycle. Assuming this number is obtained for model training, where this card can reach 624 FP16 TFLOPS \cite{nvidia_nvidia_2022}, this would still amount to ~2400 $\frac{gCO_2eq}{10^{18}FLOP}$. This study however does not include datacenter related emissions. 

\subsubsection{GPT-4 Training emissions} 

From various speculative sources based on hardware availability and compute time it can be estimated, that the computational load of GPT-4s training made up between 21 and 30 Million $10^{18}FLOP$ \cite{rahman_over_2025, mcaleese_gpt-4_2023}. While GPT-3 with its dense 175 Billion parameter model amounted to ~552 tCO2e emisions \cite{faiz_llmcarbon_2024}, GPT-4s (estimated 8x220 B size \cite{chintala_x_2023, sairjy_gpt-4_2023}) training thereby released an estimate of  4200 to 6000 tCO2eq if multiplied by a very conservative CCI value of 200 $\frac{gCO_2eq}{10^{18}FLOP}$. This corresponds to a CCI value lower then even the most optimal configuration according to Schneider et al. and towards the higher end of the estimations of d'Orgeval et al., who estimate 4 300 to 15 005 tCO2eq themselves for the comparably sized GPT-4o's Training.     
Another estimate like the one by K. Ludvigsen \cite{ludvigsen_carbon_2023} from 2023 comes out to ~12 456 tCO2e in a 90 day training cycle by just looking at the hardware and common assumptions around its usage. The main difference between the calculated value and the other two may be explained by the high carbon power estimate of 240 $\frac{gCO_2eq}{kWh}$ used for california based servers. This means as the model was most likely trained in the US, the higher end estimations of d'Orgeval et al. and K. Ludvigsen are more realistic. 
As the state of available data appears to be purposefully scarce and figures are connected to strong inaccuracies, there is no way to make a precise statement beyond an order of magnitude estimate for which 10 $ktCO2_{eq}$ will be used further on. 
These number further aligns with Falk et al. \cite{falk_et_al_more_2025}, who estimate 9.7 $ktCO2_{eq}$ for the training life-cycle of GPT-4. 

\subsubsection{The GPT-4 Life-cycle}

GPT-4 had a runtime of 25 months and 2 weeks as the main model on OpenAI's ChatGPT application \cite{forlini_two_2025} from 15th of March 2023 to 1st of May 2025. 
To make a statement on the impact of the training process on the carbon footprint of a single query, the emissions of the training process need to be equally divided on the total number of queries made. 
As for the amount of processed queries, we can again only estimate. From OpenAI's own communication we know their rough user base size over time as well as a statement that they handled $2.5 \cdot 10^9$  queries per day in July 2025 as stated by Sam Altman \cite{allen_altman_2025} (CEO of OpenAI). By assuming that these claims are correct and that the queries per day correlate linearly with the size of the user base as published by various sources over time \cite{shubham_chatgpt_2026} we can estimate that GPT-4 has been used in as much as 0.9 Trillion searches (Section \ref{sec:Calc}). Dividing the training emissions by the amount of searches gives an estimate on how much impact the need for the training has for the use of the model.  
\begin{equation}
    QCI_{embodied}= \frac{10 000 000 kg CO_{2eq}}{1400 000 000 000 000 queries} = 7 \cdot 10^{-6} \frac{gCO2eq}{query}
\end{equation}

Even though the emissions of 10 000 t $CO_2eq$ in the training and hardware acquiring process are a tremendous pollution resulting in ultimate carbon costs of  
\begin{equation}
    UCC_{training}= c_{CO2u} \cdot 10^4 tCO2eq = \$ 272 Million
\end{equation} ,with \begin{equation}\label{eq:molw}
    c_{CO2u} = \frac{M_{C}}{M_{CO2}} \cdot c_{Cu} = \frac{12}{44} \cdot c_{Cu} \approx 27 \cdot 10^{3}\frac{\$}{tCO_2eq}
\end{equation}, passed onto future generations. (Addressing eq.\ref{eq:molw}: as Archer et al. calculate with tons of fossil carbon and this paper calculates with tons of atmospheric $CO_2eq$, the number has to adjusted by the relative molecular weights.) 

At the scale of operation of ChatGPT this results in a per query value of one to two (depending on the query) orders of magnitude lower then the operational emissions. 

\subsubsection{GPT-4 Query Count}\label{sec:Calc} 

 \begin{table}[!h]
  \caption{Development of ChatGPT weekly active user base}
  \label{tab:gpt}
  \begin{tabular}{ccc}
    \toprule
 Days since GPT-4 launch & Million Weekly  & Billion Queries / day \\
  & Active Users\cite{shubham_chatgpt_2026} & \\

    \midrule
 859          & 670*                            & 2.3                       \\
 776          & 530*                            & 1.8                       \\
 759          & 500\cite{sarah_chatgpt_2025}    & 1.7                       \\
 687          & 400\cite{reuters_openais_2025}  & 1.4                       \\
 625          & 300\cite{openai_x_2024}         & 1.0                       \\
 503          & 200\cite{fried_openai_2024}     & 0.7                       \\
 138          & 100\cite{sarah_chatgpt_2025}    & 0.3                       \\
 0            & 60*                             & 0.2                       \\
  \bottomrule
\end{tabular}
\textit{*User base sizes for non specified points in time are linearly interpolated and rounded to the nearest 10 Million}
\end{table}

If during the time of $U_a = 730 \cdot 10^6$ weekly active users, $n_q = 2.5 \cdot 10^9$ daily queries are processed, this leads to $\approx 3.4 \frac{queries}{weekly \ user \cdot day}$ . Multiplying this with the respective amount of users known for a specific period, assuming linear growth between intervals and summing up the days until a new announcement of numbers leads to a final estimate of $Q \approx 1.4 \cdot 10^{12}$ queries made to GPT-4 during its runtime as the primary model supporting ChatGPT. 
It is noted, that inaccuracy is high due to the userbase size data at each point in time being given with only one significant digit. Additionally not all the ChatGPT requests in the given timeframe used GPT-4 as a model, lowering the real number of queries, while additional API queries have been and are made, increasing the figure. Narrowing this down to a more precise estimate should be subject to further investigation.  
$2.5 \cdot 10^9$ queries with an average token length of $t_q \approx 1000$ and $CCI \approx 1000 \frac{gCO_2eq}{10^{18}FLOP}$ would result in $1.1 ktCO_2eq$ of emissions or $\$30 Million$ of carbon cost by queries to ChatGPT every day. Comparing the findings in table \ref{tab:gpt} to OpenAI's actual emissions would be an important comparison to estimations in Section \ref{sec:oper}, while also the actual average query size could be better approximated. Since OpenAI is not disclosing their annual emissions or estimates on single query emissions it is not possible to do this as of now. As OpenAI is publishing 3rd party estimates on their own website \cite{openai_environmental_2025}, it can be assumed that they do not have internal estimates that contradict those findings in a significant way that would lower that energy usage estimate.

\subsection{Per Query emissions}\label{sec:oper}

When it comes to the emissions of just one singular query, we can insert the CCI as established earlier on in Section \ref{sec:CCI} and combine it with the amount of compute actually used in a single query. According to J. You \cite{you_how_2025} a single token takes about two FLOP for every active parameter in the model.  A 'token' in the context of LLMs refers to the minimal semantically distinct unit of language, that is processed by the LLM. 
\begin{quote}
    G|P|T|-|4| Token|izes| a| sentence| like| this|. \cite{openai_openai_2026}
\end{quote}
 Therefore $TC = 2 \cdot w_a$, which brings the compute per query QC to $QC = 2 \cdot w_{a} \cdot l_{t}$ ($l_{t}$ being the amount of processed tokens per query). 
To stay with GPT-4 and its alleged $w_{a} = 220 billion$ Mixture of Experts (MoE) architecture \cite{chintala_x_2023}, that would amount to \begin{equation}
     TC = r_{t} \cdot w_a  = 2 \cdot \frac{FLOP}{Token} \cdot 220 \cdot 10^{9}  = 440 \cdot 10^{9} \frac{FLOP}{Token}
\end{equation} 
 Now the amount of processed tokens per query can drastically vary. While Chiang et al. identify an average responds length of 261 tokens \cite{chiang_et_al_chatbot_2024}, agentic systems and increasingly complex services relying on LLM API access, such as CursorAI or Openclaw can easily use 50 000 Tokens to accomplish a minor task for an advanced user. As service providers try to further reduce friction and incorporate additional layers of computation, the number of computed tokens per query is rising. Such additions include features like, but not limited to: speech based input or output, agentic problem solving, 'reasoning' models, capabilities to process, manipulate or create different types of documents or the option to search the web in real time. Therefore the amount of tokens per query is going far beyond the proposed 261 tokens, as it would be expected in a simple call-and-responds type chat. 
 This again presents the need to make an estimate in a very wide range of possible scenarios. 

 \begin{table*}[h]
  \caption{Query and Token Carbon Intensity and Cost for different scenarios}
  \label{tab:calc}
  \begin{tabular}{cccccccc}
    \toprule
 $CCI / \frac{gCO_2eq}{10^{18}FLOP}$ & $l_{t} / \frac{Token}{query}$ & $QC / \frac{FLOP}{query}$ & $QCI / \frac{gCO_2eq}{query}$ & $QCC / \frac{cent}{query}$ & $TC / \frac{FLOP}{Token}$ & $TCI / \frac{gCO_2eq}{Token}$ & $TCC / \frac{\$}{Token}$ \\
    \midrule
 1000                            & $10^3$                        & $4.4 \cdot 10^{14}$       & $4.4 \cdot 10^{-1}$           & 1.2                      & $440 \cdot 10^{9}$        & $4.4 \cdot 10^{-4}$           & $1.2 \cdot 10^{-5}$      \\
1000                            & $3 \cdot 10^4$                & $1.3 \cdot 10^{16}$       & 13.2                          & 36                       & $440 \cdot 10^{9}$        & $4.4 \cdot 10^{-4}$           & $1.2 \cdot 10^{-5}$      \\
  \bottomrule
\end{tabular}
\end{table*}

A 'simple' scenario  could equal a text input of 1000 tokens, so a rather short (text only) prompt with the context of a small previous conversation (totalling roughly 2 pages of text). 
The 'complex' scenario here equals a complex problem solution including e.g. the creation of a short snippet of code from a prompt including a file based input using an agentic software totalling 30 000 tokens.
The CCI for this calculation example is estimated to be around $1000 \frac{gCO2eq}{10^{18}FLOP}$. This corresponds to the upper part of the range proposed by Schneider et al. \cite{schneider_et_al_life-cycle_2025}, who assumes TPU effciency, of hardware built to support Google AI.   
 
Popular estimates of these footprints \cite{you_how_2025, de_vries_growing_2023, luccioni_et_al_estimating_2023} are comparable to the findings listed in the table above. J. You arrives at a compute demand of  $0.1 \cdot 10^{15} \frac{FLOP}{query}$. The difference being explained by a model size of only 100 billion parameters and a token length of only 500. Using the calculation logic of J.You \cite{you_how_2025}, the $13.2 \cdot 10^{15} FLOP$ would result in the energy usage of $P_q \approx 39 Wh$, if run on an Nvidia H100 as per the given calculation example. In a hypothetical, california-based datacenter with a grid carbon Intensity (CI) of $CI_{Cal} = 0.252 \frac{gCO2eq}{Wh}$ \cite{california_air_ressources_board_2025_nodate}, the emissions would amount to  $QCI \approx 9.82 \frac{gCO2eq}{query}$, thereby further validating the findings. d'Orgeval et al. even estimate 10.2 to 34.4 $\frac{gCO2eq}{query}$ (15.4 $\frac{gCO2eq}{query}$ median) for a query of GPT4o with a substantially smaller Token count of just 100 Input and 400 Output Tokens.  
Neglecting the embodied footprint of $QCI_{embodied}= 7.1 \cdot 10^{-6} \frac{gCO2eq}{query}$ the carbon cost of a single complex query comes out to 
\begin{equation}
    QCC_{total} \approx QCC_{operational} = c_{CO2u} \cdot QCI
\end{equation}
\begin{equation}
    QCC_{total}  = 27 \cdot 10^{3}\frac{\$}{tCO_2eq} \cdot 1.3 \cdot 10^{-5} \frac{tCO2eq}{query} \approx  36 \frac{cent}{query} 
\end{equation} 
 Consequently, under the assumptions made above and connected to their uncertainty, every complex query to a LLM results in total costs of ¢36 to the future human population due to the impact of the emitted greenhouses gases alone. 

Broadly speaking it is also possible to investigate the per token cost (TCC) which then only depends on the CCI of the computing facility. For every specific CCI the Token compute intensity (TCI) and token compute cost (TCC) are as follows:

\begin{equation}
    TCC = TCI \cdot c_{CO2u} = TC \cdot CCI \cdot c_{CO2u}= r_{t} \cdot w_{a} \cdot CCI \cdot c_{CO2u}
\end{equation} 

 This results in $TCI \approx 440 \cdot 10^{-6} \frac{gCO2eq}{token}$ or $TCC \approx 1.2 \cdot 10^{-3} \frac{cent}{token}$. However, while using TCI instead of QCI improves the comparability and will be useful for improving the estimation in updated versions, it creates a new layer of abstraction and is therefore not the focus of the discussion.   

\section{Discussion} 

The calculation yields multiple findings. The ultimate cost of a ChatGPT query is around QCC $= 0.4 \frac{\$}{query}$. While this cost is significant in the sense that it is non neglectable, it appears less shocking then expected. However multiple important factors stand in contrast to that impulse. Primarily the fact, that \$0.4 are only connected to a single complex query.

With heavy usage this can pile up very quickly. Taken for example the recently leaked token usage leaderboards of Meta and OpenAI with the top Meta employee using 281 billion tokens in a month, while a OpenAI employee allegedly even spent 210 Billion tokens in one week. Using the TCI and TCC from above \ref{sec:oper} This would correspond to $120 tCO_2eq$ or $\$3.4 Million$ in a month and $92 tCO_2eq$ or $\$2.5 Million$ in a week respectively. The entirety of the Meta staff has spent $\approx$ 60 Trillion tokens \cite{munis_meta_2026} resulting in $26 400 tCO_2eq$ or a ultimate carbon cost of $\$720 Million$. These token usage counts however are immensely artificially inflated and stand in no relation to any real world benefits. Even the computation of 100 complex problems every day of the month would only lead to something around 90 Million tokens spent (30000 token per query times 100 queries times 30 days) or $40 kgCO_2eq$ and $\$1k$ carbon cost. This goes to show the absolute absurdity of token usage leaderboards and its devastating effect on our planetary health.   
Another important aspect is, that GPT-4 is only one of very many increasingly complex models. As by January 2025 over 30 models have been trained to the size of GPT-4 \cite{rahman_over_2025}, which amounts to carbon costs of ~\$8 billion (or 300 kt $CO_2eq$ ). This is equal to the annual emissions of ~33 thousand europeans \cite{eurostat_eu_2024}.  



For comparison, the average, monthly, per capita emissions of a european citizen amount to ~752kg $CO_2eq$ \cite{eurostat_eu_2024}. This is equivalent to using ~1.7 billion tokens, which is a lot more then a realistic usage. Only with the proposed very heavy usage of 90 Million tokens per month would this start to become a considerable part (~5\%) of an individuals attributed monthly emissions. Still, even a less heavy usages mark a small but noticeable increase in emissions in our timely context, where the per capita carbon emissions are supposed to be significantly decreasing \cite{shukla_et_al_climate_2022}. 

Therefore it can be concluded that the impact AI services have on a global scale will present future generations with immense burdens in environmental damages, that have to be more prominently expressed in the discussion surrounding AI. Meanwhile the individual impact of the computational, operational emissions can still be considered small compared to other common everyday activities, such as heating or commuting. In the rapid developing field of AI this must still be continuously monitored and adjusted and might change in the future or with the implementation of improved base assumptions.   

A way to make a more meaningful statement on the global economy scale impact of the AI industry could be to view the mathematical approximation from the other way around, by estimating the global greenhouse gas emissions by the AI industry sector and devise that by the amount of fulfilled queries. While this also includes operations not connected to LLMs, it offers a very complete picture. However, while there are good resources for the AI industries power usage \cite{iea_energy_2025}, there are important gaps in the availability of the fulfilled queries per company per year, their respective market share \cite{darius_ai_2026} and the connected energy mix of the grids used by the computation facilities.
Another major point of uncertainty is the actual average token length and therefore required compute of a contemporary complex AI service query that goes far beyond a call-responds scheme. Additionally it would be possible to differentiate those services further in order to create a more nuanced overview discerning the large amount of services hidden within the well known and low friction chat interfaces, contributing to literacy in their usage. This would need to be updated with the availability and usage of services.   
It might appear to be most reasonable to be conservative both in the choosing of the token usage as well as the CCI as well as he potential cost of ultimate carbon, leading to a minimum and less disputable outcome. This would however disregard, that the reality of the AI economy is not shaped around ecological choices beyond their short term economical benefit and risks vastly underestimating real conditions. As of now, the carbon reporting in the AI industry remains voluntary and incomplete \cite{halgamuge_emissions_2026}. Therefore different calculation examples are made. While staying with GPT-4 as the model of example, a 'beneficial' and 'complex' scenario is made out. 
\subsection{Limitations and Errors}
The most significant unknown factor in the calculation of the QCI the number of tokens that constitute a query. In the range from 1k to 100k tokens the equivalent carbon cost ranges from $1.2 \frac{cent}{query}$ to $120 \frac{cent}{query}$. Further research into realistic average per-query-token-counts including the differentiation of input and output token usage is needed to reduce this range.
Additional significant uncertainties to the calculated number include the CCI (Range 60 to 2400 $\frac{gCO_2eq}{10^{18}FLOP}$), the uncertainty in the model size, the compute efficiency as well as the error range of ultimate carbon cost (\$10k/t to \$750k/t), that all factor in linearly.
Aside the limitations of the calculation, there is further oversights, like the consideration, if the observed datacenter is using its access heat productively or the plethora of other impacts, that AI operation can have including, but not limited to: human toxicity, land use, localized heat emissions, acidification, eutrophication and water use. 
While this paper focuses only on greenhouse gas emissions, water scarcity does pose a significant issue \cite{li_et_al_making_2025}, which is a highly localised one. Putting a monetary value to AI's water use appears as the next significant step towards the exploration of an ultimate cost figure. 

It has to be acknowledged that the consequences of a query may not be solely attributed to the user of the service. The mitigation of the consequences of AI's costs needs to be addressed the way it is created - on all levels from individual to systemic.    

\section{AI Statement}
Generative AI has not been used in the creation of this paper.

\bibliographystyle{ACM-Reference-Format}
\bibliography{bibliography}

@article{faiz_llmcarbon_2024,
	title = {{LLMCARBON}: Modeling the End-End carbon footprint of large language models},
	author = {Faiz, Ahmad and Kaneda, Sotaro and Wang, Ruhan and Osi, Rita and Sharma, Prateek and Chen, Fan and Jiang, Lei},
	date = {2024},
	langid = {english},
}

@online{sairjy_gpt-4_2023,
	title = {{GPT}-4 comment section},
	url = {https://www.lesswrong.com/posts/pckLdSgYWJ38NBFf8/gpt-4?commentId=2mKqGJnf2aTfQMZDq},
	author = {{sairjy}},
	urldate = {2026-05-28},
	date = {2023-03-14},
}

@online{openai_x_2024,
	title = {X Post by {OpenAI}},
	url = {https://x.com/OpenAINewsroom/status/1864373399218475440},
	author = {{OpenAI}},
	date = {2024-12-04},
}

@article{luccioni_et_al_estimating_2023,
	title = {Estimating the Carbon Footprint of {BLOOM},  a 176B Parameter Language Model},
	author = {{Luccioni et al.}},
	date = {2023-05},
	langid = {english},
}

@online{fried_openai_2024,
	title = {{OpenAI} says {ChatGPT} usage has doubled since last year},
	url = {https://www.axios.com/2024/08/29/openai-chatgpt-200-million-weekly-active-users},
	author = {Fried, Ina},
	date = {2024-08-29},
}

@report{iea_energy_2025,
	title = {Energy and {AI}},
	url = {https://www.iea.org/reports/energy-and-ai},
	author = {{IEA}},
	date = {2025},
}

@online{reuters_openais_2025,
	title = {{OpenAI}'s weekly active users surpass 400 million},
	url = {https://www.reuters.com/technology/artificial-intelligence/openais-weekly-active-users-surpass-400-million-2025-02-20/},
	author = {{Reuters}},
	date = {2025-02-20},
}

@online{darius_ai_2026,
	title = {{AI} Market Share By Company Statistics 2026},
	url = {https://www.companieshistory.com/ai-market-share-by-company/},
	author = {{Darius}},
	date = {2026-05-23},
}

@online{rahman_over_2025,
	title = {Over 30 {AI} models have been trained at the scale of {GPT}-4},
	url = {https://epoch.ai/data-insights/models-over-1e25-flop},
	author = {Rahman, Robi and Heindrich, Lovis},
	urldate = {2026-05-28},
	date = {2025-01-30},
}

@article{archer_et_al_ultimate_2020,
	title = {The ultimate cost of carbon},
	volume = {162},
	issn = {0165-0009, 1573-1480},
	url = {https://link.springer.com/10.1007/s10584-020-02785-4},
	doi = {10.1007/s10584-020-02785-4},
	pages = {2069--2086},
	number = {4},
	journaltitle = {Climatic Change},
	shortjournal = {Climatic Change},
	author = {{Archer et al.}},
	urldate = {2026-06-10},
	date = {2020-10},
	langid = {english},
}

@article{devadas_towards_2026,
	title = {Towards carbon-aware {AI}: a systematic prisma review and taxonomy of green architectures, hardware life-cycle, and energy-efficient algorithms},
	volume = {9},
	issn = {2520-8942},
	url = {https://link.springer.com/10.1186/s42162-026-00651-8},
	doi = {10.1186/s42162-026-00651-8},
	shorttitle = {Towards carbon-aware {AI}},
	pages = {38},
	number = {1},
	journaltitle = {Energy Informatics},
	shortjournal = {Energy Inform},
	author = {Devadas, Raghavendra M. and T, Sowmya},
	urldate = {2026-05-14},
	date = {2026-03-25},
	langid = {english},
}

@online{sarah_chatgpt_2025,
	title = {{ChatGPT} doubled its weekly active users in under 6 months, thanks to new releases},
	url = {https://techcrunch.com/2025/03/06/chatgpt-doubled-its-weekly-active-users-in-under-6-months-thanks-to-new-releases/},
	author = {Sarah, Perez},
	date = {2025-03-06},
}

@article{schneider_et_al_life-cycle_2025,
	title = {Life-Cycle Emissions of {AI} Hardware: A Cradle-To-Grave Approach and Generational Trends},
	author = {{Schneider et al.}},
	date = {2025},
	langid = {english},
}

@article{brennan_considering_2017,
	title = {Considering the public private-dichotomy: Hannah Arendt, Václav Havel and Victor Klemperer on the importance of the privat},
	volume = {40},
	issn = {0163-8548, 1572-851X},
	url = {http://link.springer.com/10.1007/s10746-017-9424-x},
	doi = {10.1007/s10746-017-9424-x},
	shorttitle = {Considering the Public Private-Dichotomy},
	pages = {249--265},
	number = {2},
	journaltitle = {Human Studies},
	shortjournal = {Hum Stud},
	author = {Brennan, Daniel},
	urldate = {2026-06-09},
	date = {2017-06},
	langid = {english},
}

@misc{li_et_al_making_2025,
	title = {Making {AI} Less "Thirsty"},
	url = {http://arxiv.org/abs/2304.03271},
	doi = {10.48550/arXiv.2304.03271},
	shorttitle = {Making {AI} Less "Thirsty"},
	number = {{arXiv}:2304.03271},
	publisher = {{arXiv}},
	author = {{Li et al.}},
	urldate = {2026-05-15},
	date = {2025-03-26},
	langid = {english},
	eprinttype = {arxiv},
	eprint = {2304.03271 [cs.LG]},
}

@collection{shukla_et_al_climate_2022,
	location = {Geneva},
	title = {Climate Change 2022 Mitigation of Climate Change Summary for Policymakers},
	isbn = {978-92-9169-160-9},
	shorttitle = {Climate change 2022},
	pagetotal = {1},
	publisher = {{IPCC}},
	editor = {{Shukla et al.} and {IPCC}},
	date = {2022},
}

@online{you_how_2025,
	title = {How much energy does {ChatGPT} use?},
	url = {https://epoch.ai/gradient-updates/how-much-energy-does-chatgpt-use},
	author = {You, Josh},
	date = {2025-02-07},
}

@online{forlini_two_2025,
	title = {Two Years After {GPT}-4 Broke the Internet, {OpenAI} Is Quietly Killing It},
	url = {https://uk.pcmag.com/ai/157540/two-years-after-gpt-4-broke-the-internet-openai-is-quietly-killing-it},
	author = {Forlini, Emily},
	urldate = {2026-05-28},
	date = {2025-04-11},
}

@online{chintala_x_2023,
	title = {X Post},
	url = {https://x.com/soumithchintala/status/1671267150101721090},
	author = {Chintala, Soumith},
	urldate = {2026-05-28},
	date = {2023-06-20},
}

@online{mcaleese_gpt-4_2023,
	title = {{GPT}-4 Predictions},
	url = {https://www.lesswrong.com/posts/qdStMFDMrWAnTqNWL/gpt-4-predictions},
	author = {{McAleese}, Stephen},
	urldate = {2026-05-28},
	date = {2023-02-18},
}

@misc{eurostat_eu_2024,
	title = {{EU} {GHG} emissions from the production and consumption (footprint) perspectives ({FIGARO} application)},
	url = {https://ec.europa.eu/eurostat/databrowser/product/page/CLI_GGE_FOOT},
	doi = {10.2908/CLI_GGE_FOOT},
	publisher = {Eurostat},
	author = {{Eurostat}},
	urldate = {2026-06-10},
	date = {2024},
}

@article{halgamuge_emissions_2026,
	title = {Emissions transparency in {AI} research should be mandatory},
	volume = {50},
	issn = {22105379},
	url = {https://linkinghub.elsevier.com/retrieve/pii/S2210537926000235},
	doi = {10.1016/j.suscom.2026.101313},
	pages = {101313},
	journaltitle = {Sustainable Computing: Informatics and Systems},
	shortjournal = {Sustainable Computing: Informatics and Systems},
	author = {Halgamuge, Malka N. and Srinivasa, Narayan},
	urldate = {2026-05-14},
	date = {2026-06},
	langid = {english},
}

@article{dorgeval_et_al_generative_2026,
	title = {Generative {AI} impact assessment through a life cycle analysis of multiple data center typologies},
	volume = {406},
	issn = {03062619},
	url = {https://linkinghub.elsevier.com/retrieve/pii/S0306261925020185},
	doi = {10.1016/j.apenergy.2025.127288},
	pages = {127288},
	journaltitle = {Applied Energy},
	shortjournal = {Applied Energy},
	author = {{d'Orgeval et al.}},
	urldate = {2026-05-14},
	date = {2026-03},
	langid = {english},
}

@article{de_vries_growing_2023,
	title = {The growing energy footprint of artificial intelligence},
	volume = {7},
	issn = {25424351},
	url = {https://linkinghub.elsevier.com/retrieve/pii/S2542435123003653},
	doi = {10.1016/j.joule.2023.09.004},
	pages = {2191--2194},
	number = {10},
	journaltitle = {Joule},
	shortjournal = {Joule},
	author = {De Vries, Alex},
	urldate = {2026-06-08},
	date = {2023-10},
	langid = {english},
}

@online{shubham_chatgpt_2026,
	title = {{ChatGPT} Statistics (June 2026) – Latest Active Users Data},
	url = {https://www.demandsage.com/chatgpt-statistics/},
	author = {Shubham, Singh},
	urldate = {2026-06-06},
	date = {2026-05-06},
}

@misc{chiang_et_al_chatbot_2024,
	title = {Chatbot Arena: An Open Platform for Evaluating {LLMs} by Human Preference},
	rights = {{arXiv}.org perpetual, non-exclusive license},
	url = {https://arxiv.org/abs/2403.04132},
	doi = {10.48550/ARXIV.2403.04132},
	shorttitle = {Chatbot Arena},
	publisher = {{arXiv}},
	author = {{Chiang et al.}},
	urldate = {2026-06-08},
	date = {2024},
	note = {Version Number: 1},
}

@online{ludvigsen_carbon_2023,
	title = {The carbon footprint of {GPT}-4},
	url = {https://towardsdatascience.com/the-carbon-footprint-of-gpt-4-d6c676eb21ae/},
	author = {Ludvigsen, Kasper Groes Albin},
	urldate = {2026-05-28},
	date = {2023-07-18},
}

@article{jacquet_et_al_cinergy_2026,
	title = {{CINERGY}: Deterministic Power Monitoring for  Carbon Accounting in the Cloud},
	rights = {https://ieeexplore.ieee.org/Xplorehelp/downloads/license-information/{IEEE}.html},
	issn = {2168-7161, 2372-0018},
	url = {https://ieeexplore.ieee.org/document/11434992/},
	doi = {10.1109/TCC.2026.3674370},
	shorttitle = {Cinergy},
	pages = {1--12},
	journaltitle = {{IEEE} Transactions on Cloud Computing},
	shortjournal = {{IEEE} Trans. Cloud Comput.},
	author = {{Jacquet et al.}},
	urldate = {2026-05-14},
	date = {2026},
	langid = {english},
}

@article{california_air_ressources_board_2025_nodate,
	title = {2025 Carbon Intensity Values for California Average Grid Electricity},
	journaltitle = {24/03/2025},
	author = {California Air Ressources Board},
	langid = {english},
}

@online{allen_altman_2025,
	title = {Altman plans D.C. push to "democratize" {AI} economic benefits},
	url = {https://www.axios.com/2025/07/21/sam-altman-openai-trump-dc-fed},
	author = {Allen, Mike},
	urldate = {2026-05-28},
	date = {2025-07-21},
}

@article{pinski_ai_2024,
	title = {{AI} literacy for users – A comprehensive review and future research directions of learning methods, components, and effects},
	volume = {2},
	issn = {29498821},
	url = {https://linkinghub.elsevier.com/retrieve/pii/S2949882124000227},
	doi = {10.1016/j.chbah.2024.100062},
	pages = {100062},
	number = {1},
	journaltitle = {Computers in Human Behavior: Artificial Humans},
	shortjournal = {Computers in Human Behavior: Artificial Humans},
	author = {Pinski, Marc and Benlian, Alexander},
	urldate = {2026-06-09},
	date = {2024-01},
	langid = {english},
}

@article{kirkpatrick_carbon_2023,
	title = {The Carbon Footprint of Artificial Intelligence},
	volume = {66},
	issn = {0001-0782, 1557-7317},
	url = {https://dl.acm.org/doi/10.1145/3603746},
	doi = {10.1145/3603746},
	pages = {17--19},
	number = {8},
	journaltitle = {Communications of the {ACM}},
	shortjournal = {Commun. {ACM}},
	author = {Kirkpatrick, Keith},
	urldate = {2026-05-14},
	date = {2023-08},
	langid = {english},
}

@book{crawford_atlas_2021,
	location = {New Haven London},
	title = {Atlas of {AI}},
	isbn = {978-0-300-26463-0 978-0-300-25239-2},
	shorttitle = {Atlas of {AI}},
	pagetotal = {1},
	publisher = {Yale University Press},
	author = {Crawford, Kate},
	date = {2021},
}

@online{openai_environmental_2025,
	title = {Environmental Impact of {AI}},
	url = {https://academy.openai.com/public/clubs/higher-education-05x4z/resources/environmental-impact-of-ai},
	author = {{OpenAI}},
	urldate = {2026-06-30},
	date = {2025-08-22},
}

@software{openai_openai_2026,
	title = {{OpenAI} Tokenizer},
	url = {https://platform.openai.com/tokenizer},
	author = {{OpenAI}},
	urldate = {2026-06-29},
	date = {2026-06-29},
}

@inproceedings{bender_et_al_dangers_2021,
	location = {Virtual Event Canada},
	title = {On the Dangers of Stochastic Parrots: Can Language Models Be Too Big?},
	isbn = {978-1-4503-8309-7},
	url = {https://dl.acm.org/doi/10.1145/3442188.3445922},
	doi = {10.1145/3442188.3445922},
	shorttitle = {On the Dangers of Stochastic Parrots},
	eventtitle = {{FAccT} '21: 2021 {ACM} Conference on Fairness, Accountability, and Transparency},
	pages = {610--623},
	booktitle = {Proceedings of the 2021 {ACM} Conference on Fairness, Accountability, and Transparency},
	publisher = {{ACM}},
	author = {{Bender et al.}},
	urldate = {2026-06-29},
	date = {2021-03-03},
	langid = {english},
}

@online{munis_meta_2026,
	title = {A Meta employee created a dashboard so coworkers can compete to be the company’s No. 1 {AI} token user},
	url = {https://fortune.com/2026/04/09/meta-killed-employee-ai-token-dashboard/},
	author = {Munis, Jacqueline},
	urldate = {2026-06-28},
	date = {2026-09-04},
}

@misc{falk_et_al_more_2025,
	title = {More than Carbon: Cradle-to-Grave environmental impacts of {GenAI} training on the Nvidia A100 {GPU}},
	url = {http://arxiv.org/abs/2509.00093},
	doi = {10.48550/arXiv.2509.00093},
	shorttitle = {More than Carbon},
	number = {{arXiv}:2509.00093},
	publisher = {{arXiv}},
	author = {{Falk et al.}},
	urldate = {2026-08-11},
	date = {2025-12-18},
	langid = {english},
	eprinttype = {arxiv},
	eprint = {2509.00093 [cs.CY]},
}

@article{nvidia_nvidia_2022,
	title = {{NVIDIA} A100 {\textbar} Tensor Core {GPU}},
	author = {{NVIDIA}},
	date = {2022},
	langid = {english},
}


\appendix


  

\end{document}